\documentclass[twocolumn,english,aps,article,showkeys,showpacs]{revtex4}
\usepackage[T1]{fontenc}
\usepackage[latin1]{inputenc}
\usepackage{babel}
\usepackage{graphics}
\usepackage{setspace}

\makeatletter

\providecommand{\LyX}{L\kern-.1667em\lower.25em\hbox{Y}\kern-.125emX\@}

\usepackage[T1]{fontenc}
\usepackage[latin1]{inputenc}
\usepackage{babel}
\usepackage{graphics}

\makeatletter

\usepackage[T1]{fontenc}
\usepackage[latin1]{inputenc}
\usepackage{babel}

\makeatletter

\usepackage[T1]{fontenc}
\usepackage[latin1]{inputenc}
\usepackage{babel}



\makeatother

\makeatother
\begin{document}

\title{Variable Range Hopping Conduction in Complex Systems and a Percolation Model with Tunneling}

\author{Asok K. Sen\footnote{Corresponding author. Fax: (+91 33) 2337 4637}} 
\email{asok@cmp.saha.ernet.in (old); asokk.sen@saha.ac.in (new)} 

\author{Somnath Bhattacharya} 
\email{somnath@cmp.saha.ernet.in (old); somnath.bhattacharya@saha.ac.in (new)}

\affiliation{Condensed Matter Physics (CMP) Group, Saha Institute of Nuclear Physics, 1/AF, Bidhan Nagar, Kolkata 700 064, India}



\vskip 1.0cm


\begin{abstract}
\noindent
For the low-temperature electrical conductance of a disordered {\it quantum
insulator} in $d$-dimensions, Mott \cite{mott} had proposed his Variable
Range Hopping (VRH) formula, $G(T) = G_0~{\rm exp}[-(T_0/T)^{\gamma}]$,
where $G_0$ is a material constant and $T_0$ is a characteristic
temperature scale.  For disordered but non-interacting carrier charges, Mott
had found that $\gamma = 1/(d+1)$ in $d$-dimensions.  Later on, Efros
and Shkolvskii \cite{esh} found that for a pure ({\it i.e.}, disorder-free)
{\it quantum insulator} with interacting charges, $\gamma =1/2$,
{\it independent of d}.  Recent experiments indicate that $\gamma$ 
is either (i) larger than any of the above predictions; and,
(ii) more intriguingly, it seems to be a function of $p$, the dopant
concentration.  We investigate this issue with a {\it semi-classical} or
{\it semi-quantum} RRTN ({\it Random Resistor cum Tunneling-bond Network})
model, developed by us in the 1990's.  These macroscopic {\it granular/
percolative composites} are built up from randomly placed meso- or
nanoscopic coarse-grained clusters, with two phenomenological functions
for the temperature-dependence of the metallic and the semi-conducting
bonds.  We find that our RRTN model (in 2D, for simplicity) also captures
this continuous change of $\gamma$ with $p$, satisfactorily.
\end{abstract}

\maketitle
\noindent The low-temperature {\it dc} electrical conductance $G(T)$ is being
studied for many decades now, in the regime where the thermal energy 
$k_BT$ ($k_B$ = Boltzmann constant) is of the order of or smaller than 
the disorder or the Coulomb interaction energy between the charge carrying 
fermions.  In the decade of 1960's, Mott \cite{mott} had put forward an 
analytical expression for the {\it phonon-assisted hopping} conduction of 
{\it spinless fermions}, taking {\it only the lattice disorder} effect into 
account and his {\it Variable Range Hopping} (VRH) formula is written as,
\begin{equation}
G(T) =  G_0~{\rm exp}\left[- \left(\frac{T_0}{T}\right)^\gamma\right],
\end{equation}
where $G_0$ is a material parameter, $\gamma= 1/(d+1)$ for a
$d$-dimensional sample (e.g., $\gamma =1/4$ in 3D), and $T_0$
is a sample-specific temperature scale, below which {\it quantum mechanical
tunneling} between nearby fermionic states (electron or hole), {\it
localized around a finite number of lattice sites}, starts contributing 
significantly to the $G(T)$ with the help of {\it hopping} due to phonons.  
Classically, these regions behave as {\it finite size clusters}.  For a 
quantum insulator, as $T$ tends to $T_0$, the {\it coherent tunneling} process 
(or, hopping conduction) keeps increasing, while the {\it incoherent 
scattering} due to the phonons (or, the ohmic resistance) keeps decreasing.  
After Mott's seminal work, Efros and Shklovskii \cite{esh} considered the 
localization due {\it only} to the {\it repulsive Coulomb interaction} between 
the charge carriers in a pure system, and achieved the complementary
result that $\gamma = 1/2$ for an insulating sample in any {\it d}.  Musing 
over both the types of VRH, one may take $k_{B}T_{0}$ as the energy-scale
 which determines the domain above which incoherent (dephasing) scattering 
among the localised electron/hole states {\it completely takes over}, and  
transforms an Anderson or a Mott insulator into an Ohmic (diffusive) metal. 
Thus, for a complete description, the VRH formula should take the following 
{\it general} form:
\begin{equation}
G(T) =  G_0~\left(\frac{T_0}{T}\right)^s {\rm exp}\left[-
\left(\frac{T_0}{T}\right)^\gamma\right],
\end{equation}
as argued by Aharony {\it et.al.} \cite{ahe}.

\noindent But, relatively recently, there have been a few theoretical works
({\it e.g.}, ref. \cite{dls}) and some experiments ({\it e.g.}, \cite{pmbbm,
ryymsh,ogas}) which do not seem to fall into any of the above schemes,
in the sense that the exponent $\gamma$ is different from the above
predictions.  While the theory of Ref.\cite{dls} predicts $\gamma >
1/4$ in 3D disordered systems on fractal media (due to hopping between
{\it superlocalized} states), and the experiment of Ref.\cite{pmbbm} on
Carbon black-PVC composites seem to confirm such hoppings in the presence of
both disorder and interaction.  Here {\it superlocalized states} are those 
whose wave-functions decay with the distance $R$ as ${\rm exp}[-(R/\xi)^{\zeta}]$, with $\zeta > 1$; $\xi$ being the localization length.  If hopping takes 
place between superlocalized states, then the Mott VRH was shown to modify the 
exponent $\gamma$ in Eq.(1) to $\gamma= d_f/(d_f+\zeta)$, where $d_f$ is the 
fractal dimension of the medium \cite{dls}.  Experimental \cite{pmbbm} evidence 
of the above has been reported in carbon-black-polymer composites, where it
is claimed that $\zeta = 1.94 \pm 0.06$.  However, doubt has been cast
by Aharony {\it et al.} \cite{ahe} whether the superlocalization was really
observed in such composites.
\vskip 0.1 in 
\noindent As a matter of fact, in the last decade, several experiments ({\it e.g.}, Ref. \cite{ryymsh,ogas}) had reported deviations from the above-mentioned results.  
Indeed, the more serious deviation has been the continuous variation of the 
exponent $\gamma$ with the dopant (or, disorder) concentration, $p$, in some 
{\it granular} or {\it composite} materials.  To capture the basic physics of
this intriguing behaviour, we undertook a thorough study of the low-temperature 
conductance in a semi-classical percolation model, called the RRTN ({\it Random Resistor cum Tunneling-bond Network}) model.  The details of the model and some 
of its applications are discussed in the original papers by Sen and co-workers 
\cite {asag1, asag2, asag3, asag4}.  Briefly stated, in the RRTN, over and 
above the randomly placed ohmic ($o$) bonds in an insulating background, 
(called a {\it Random Resistor Network}, or RRN), a {\it semi-classical tunneling} is allowed {\it only} through the {\it nearest-neighbor insulators} between 
any two $o$-bonds.  Thus, these particular insulating bonds are called tunneling
($t$) bonds.  The placement of these $t$-bonds in a deterministic way (for a 
given random resistor configuration) makes our RRTN model partly deterministic 
(statistically correlated).  It may be noted that in the RRTN model, (i) the 
\begin{figure}
\resizebox*{7cm}{7cm}{\rotatebox{270}{\includegraphics{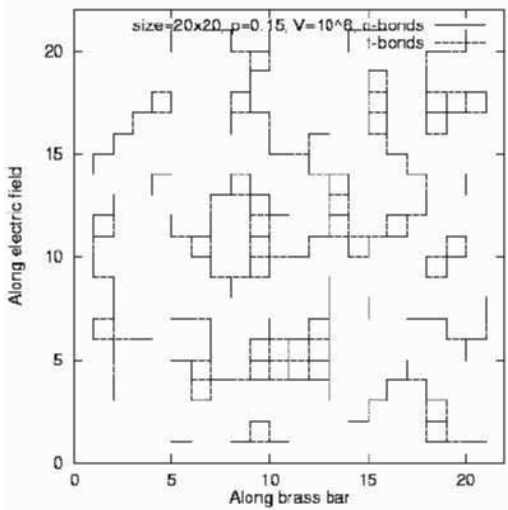}}}
\caption{An example of a 20x20 square-lattice configuration in RRTN model, at a random $o$-bond (full lines) concentration of $p$=0.15 and with at an arbitrarily high voltage of $10^8 V$, just to make sure that all the t-bonds (dashed lines, with a threshold of $v_g = 0.5 V$) are active (with an {\it inessential} assumption of no Joule heating). One may note here that both the RRN and the RRTN sample is non-percolating (both an insulator).}  
\end{figure}
\begin{figure}
\resizebox*{7cm}{7cm}{\rotatebox{270}{\includegraphics{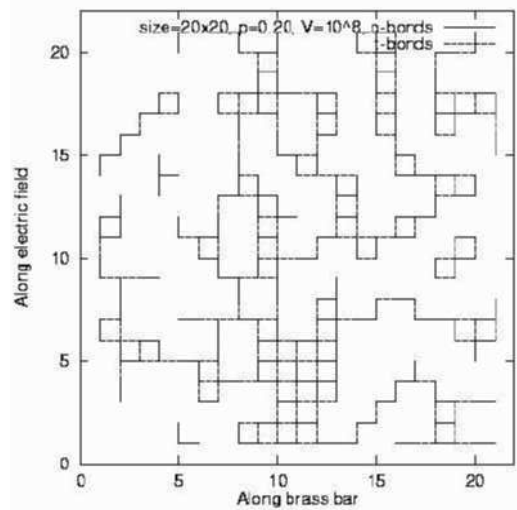}}}
\caption{Another example of a 20x20 square lattice configuration (where all the possible $t$-bonds are driven active) but at a slightly larger $p$=0.20.  Some parts of both the lattices look identical, because their construction started with the same iseed (a trivial issue).  One may observe here that the RRTN is percolating (metal) but the RRN is not (insulator).  These two Figs. 1 and 2 thus indicate a {\it new} insulator-metal transition with a percolation threshold inside $0.15 < p < 0.20$.}
\end{figure}
{\it disorder-}parameter is introduced through the random occupations of the 
$o$-bonds on the lattice, and (ii) the {\it Coulomb interaction} is mimicked 
through the {\it gap} or the threshold voltage $v_g$ for tunneling past the 
$t$-bonds.  Thus, for a given macroscopic external voltage $V$, a given $t$-bond
 becomes {\it active} ({\it i.e.}, lets tunneling take place) if the 
microscopic voltage difference across its ends is above $v_g$, and remains 
inactive if this voltage difference is below $v_g$.  We call the percolation 
threshold of the RRTN model, when all the possible $t$-bonds are active, 
$p_{ct}$.  For the RRTN embedded in a 2D-square lattice, finite-size analysis
gives $p_{ct} \simeq 0.181$ (in an Effective Medium Approximation (EMA), 
$p_{ct}$ is $0.25$) \cite{asag1, asag2}.
\vskip 0.1 in 
\noindent It may be noted that the temperature $T$ does not explicitly appear in any 
percolation model.  So, to study the VRH phenomenon in the RRTN (in a 2D-square lattice), we use some {\it empirical parameters} for the $T$-dependence of the 
{\it microscopic} conductance of the various types of bonds.  For simplicity, 
for the $o$-bonds we use $g_o = 1/(r_o+a~T)$ and for the t-bonds, $g_t = b~{\rm exp}(-c/T)$.  To go beyond the EMA, {\it i.e.}, to study the effects of both 
the thermal and the (semi)-quantum fluctuations, we had to take recourse to a 
numerical solution of the Kirchoff's laws ({\it local current conservation}) at 
each node of the lattice.  The aim is to achieve {\it global current} (which may be zero if the macroscopic sample is 
an insulator) {\it conservation} from the local ones in the RRTN.  In a 
classically percolating situation at a low voltage $V$, where the ohmic backbone
 is already percolating, we find a {\it non-monotonic} sharp rise in the 
conductance as $T \rightarrow 0 $; eventuallly saturating to a large but finite 
{\it residual resistance} $R_0$, {\it i.e.}, $R(T=0;RRTN)$, in any finite-size 
RRTN sample.  Further, $R_0$ is found to be {\it sensitively dependent on
specific configurations} ({\it i.e.}, sample-specific), for any given
concentration $p$ of the $o$-bonds.  
\begin{figure}
\resizebox*{8cm}{7cm}{\rotatebox{270}{\includegraphics{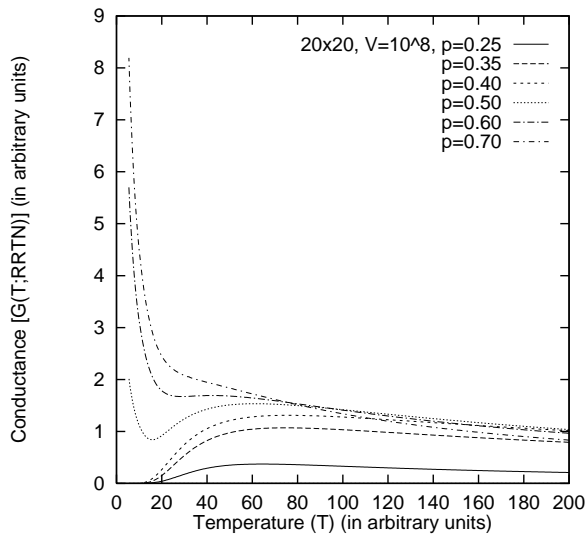}}}
\caption{Conductance $G(T)$ {\it .vs.} temperature $T$ in six 20x20 RRTN square-lattices for various $p$'s in the upper ohmic regime.  Due to finite-size effects, some are percolating and some are not.} 
\end{figure}
\begin{figure}
\resizebox*{8cm}{7cm}{\rotatebox{270}{\includegraphics{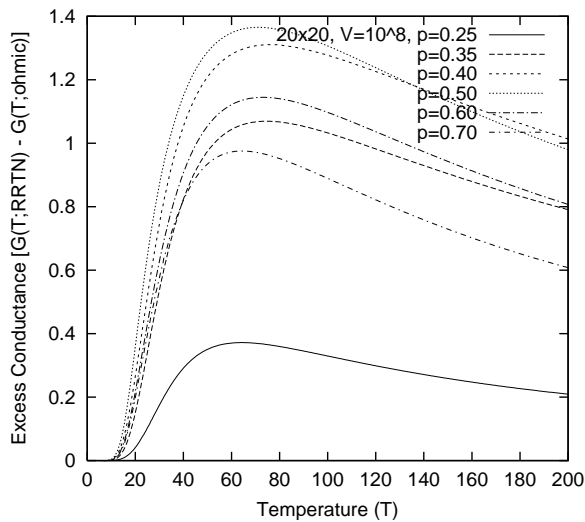}}} 
\caption{Excess $G(T)$ in the same RRTN samples, as in the Fig.3.  The word {\it excess} means that the ohmic (RRN) part of the $G(T)$ for the percolating RRTN samples at each $T$ is subtracted out.}
\end{figure}
These findings have, at least, a qualitative matching, with some experimental results on {\it composite/ granular materials}, as well as those in {\it fully quantum disordered} systems 
({\it e.g.}, see Ref. \cite{meir}).  In the Fig.3, we present the $T$-dependence of the conductance in finite-sized ($L= 20$) RRTN samples for six different 
$p$'s.  Fig.4 shows the excess effect of the $t$-bonds on the same footing for 
all the samples (both the percolating and the nonpercolating RRTN) of Fig.3,
by subtracting their RRN (ohmic) contributions at each $T$.  We calculated the 
exponents using the {\it finite size scaling} analysis.  In the Fig.5, we show 
a typical finite size effect on the 2D-RRTN in the {\it flat upper ohmic regime}
 (where all the possible $t$-bonds are active).  In our previous study \cite{kgs-temp} in this regime, the finite-size effects had seemed vanishing.  More 
precise analysis with four $L$'s (see Fig.5 for $T < T_m$) indicate that 
$\gamma \rightarrow 1$ sensitively, but $s$ is very robust ($2.0 < s < 3.0$) as 
$T \rightarrow 0$.  Also, this size-effect is stronger in the strongly nonlinear, sigmoidal regime, of the {\it dc} current-voltage response \cite{asag1,asag4}.
These details will be discussed elsewhere.
\vskip 0.1 in 
\noindent To summarize, we find that our RRTN model is quite successful in 
describing the VRH and that the {\it generalized} VRH formula, Eq.(2), works 
better than the restricted one, {\it i.e.}, Eq.(1).  To wit, the exponents 
vary continuously with the concentration $p$ of the $o$-bonds. So, it is 
capable of reproducing, at least qualitatively, the intriguing results of 
some recent experiments on 
(Carbon-black)-(PVC) mixture \cite{pmbbm}, sulfonated polyaniline (called PANI-CSA) composites \cite{ryymsh}, some doped Langmuir-Blodgett films \cite{ogas} 
etc.  Of course, the EMA-values of $s$ and $\gamma$ differ from their respective numerically "exact" values; {\it e.g.}, in the EMA, $s=0.5$ for all the $p$'s. 
Further theoretical studies would possibly involve the {\it non-extensivity and some generalized entropy}, defined as a {\it measure of loss of information}, 
as formulated by Tsallis' and Re'nyi (see, {\it e.g.}, ref.\cite{abe} by Abe), 
for the far-from-equilibrium dynamics of the RRTN.
\begin{figure}
\resizebox*{8cm}{7cm}{\rotatebox{270}{\includegraphics{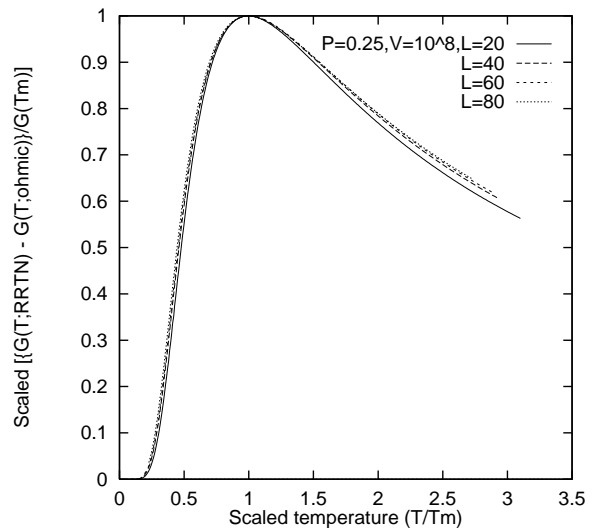} }}  
\caption{Normalized VRH conductances in the RRTN model at $p=0.25$ for 2D-square lattices of various sizes, in the limit where all the allowed $t$-bonds are active.  The maxima values $G_m=G(T_m, L)$ and the corresponding  $T_m=T_m(L, p)$'s are used to normalize each curve.  The fittings of the VRH parameters for each $L$ use the full curves.  The finite-size effects are {\it NOT too prominent} for $T < T_m$.}
\end{figure}
\vskip 0.5cm
\begin{acknowledgments}
\noindent AKS would like to thank Prof. D.J. Bergman for very warm hospitality 
and support for presenting this work at the International Symposium on
"Continuum Models and Discrete Systems" (CMDS10) held in Shoresh, Israel,
during 30 June - 04 July, 2003.  During this activity, AKS had useful
discussions with Profs. N. Argaman, H. Herrmann and H. Levine.
AKS would also like to thank the organizers of "High Level
European Conferences," for their support to present an earlier version of
the present work in the 2nd. European Summer School on "Microscopic
Quantum Many Body Theories and their Applications", held at the Abdus Salam
International Centre for Theoretical Physics (AS-ICTP), Trieste, Italy,
during 03-14 September, 2001.  SB would like to thank the AS-ICTP and
Prof. V.E. Kravtsov (resident convenor), for the financial support to
participate with a short talk on variable range hopping in our RRTN model,
at a Workshop on "Mesoscopic Physics and Electron Interactions" held at
the AS-ICTP during 24 June - 05th. July 2002; and for some useful
discussions with him and with Prof. K.B. Efetov (Ref.\cite{efetov}).
\end{acknowledgments}


\begin{thebibliography}{1}

\bibitem{mott} N.F. Mott, J. Non-Cryst. Solids {\bf 1}, 1 (1968); V.
Ambegaokar, B.L. Halperin and J.S. Langer, Phys. Rev. B {\bf 4}, 2612 (1971)

\bibitem{esh} A.L. Efros and B.I. Shklovskii, J. Phys. C {\bf 8}, 249 (1975)

\bibitem{ahe} A. Aharony, A.B. Harris, and O. Entin-Wohlman, Phys. Rev.
Lett. {\bf 70}, 4160 (1993)

\bibitem{dls} G. Deutscher, Y.-E. L\'{e}vy, and B. Souillard, Europhys.
Lett. {\bf 4}, 577 (1987)

\bibitem{pmbbm} D. van der Putten, J.T. Moonen, H.B.Brom, J.C.M.
Brokken-Zijp, and M.A.J. Michels, Phys. Rev. Lett. {\bf 69}, 494 (1992)

\bibitem{ryymsh} M. Reghu, C.O. Yoon, C.Y. Yang, D. Moses, P. Smith, and
A.J. Heeger, Phys. Rev. B {\bf 50}, 13931 (1994)

\bibitem{ogas} K. Ogasawara, T. Ishiguro, S. Horiuchi, H. Yamochi, G.
Saito, and Y. Nogami, J. Phys. Chem. Solids {\bf 58}, 39 (1997)

\bibitem{asag1} \emph {Scaling of Electric Response in a Non-linear Composite}
Asok K.  Sen and A. Kar Gupta) - Proceedings of an Indo-French Workshop on
\emph{Non-linearity and Breakdown in Soft-Condensed Matter}, held at the Saha
Institute of Nuclear Physics, Calcutta, 1993, Lecture Notes in Physics Series
v.{\bf 437}, p.271-287, Eds. K.K. Bardhan, B.K. Chakrabarti and A. Hansen,
(Springer-Verlag, Berlin, 1994).

\bibitem{asag2} A. Kar Gupta and Asok K. Sen, Physica A {\bf 215}, 1 (1995)

\bibitem{asag3} A. Kar Gupta and Asok K. Sen, Physica A {\bf 247}, 30 (1997)

\bibitem{asag4} A. Kar Gupta and Asok K. Sen, Phys. Rev. B {\bf 57},
3375 (1998)

\bibitem{kgs-temp} A preliminary study of the VRH in the RRTN was published
by: A. Kar Gupta, D. Dan and A.K. Sen, Ind. J. Phys. {\bf 71A}, 357 (1997)

\bibitem{meir} Y. Meir, Phys. Rev. B {\bf 61}, 16470 (2000).

\bibitem{abe} S. Abe, Physica A {\bf 300}, 417 (2001)

\bibitem{efetov} K.B. Efetov and A. Tschersich; arXiv: cond-matt/0109033 (2001)

\end{thebibliography}
\end{document}